# VOFilter, Bridging Virtual Observatory and Industrial Office Applications

Chen-zhou Cui[1], Markus Dolensky[2], Peter Quinn[2], Yong-heng Zhao[1] and Françoise Genova[3]

[1]National Astronomical Observatories, Chinese Academy of Sciences, Beijing 100012; ccz@bao.ac.cn
[2]European Southern Observatory, D-85748 Garching bei München, Germany
[3]CDS, Observatoire Astronomique de Strasbourg, France

**Abstract**

VOFilter is an XML based filter developed by the Chinese Virtual Observatory project to transform tabular data files from VOTable format into OpenDocument format. VOTable is an XML format defined for the exchange of tabular data in the context of the Virtual Observatory (VO). It is the first Proposed Recommendation defined by International Virtual Observatory Alliance, and has obtained wide support from both the VO community and many Astronomy projects. OpenOffice.org is a mature, open source, front office applications suite with the advantage of native support of industrial standard OpenDocument XML file format. Using the VOFilter, VOTable files can be loaded in OpenOffice.org Calc, a spreadsheet application, and then displayed and analyzed as other spreadsheet files. Here, the VOFilter acts as a connector, bridging the coming VO with current industrial office applications. Virtual Observatory and technical background of the VOFilter are introduced. Its workflow, installation and usage are presented. Existing problems and limitations are also discussed together with the future development plans.

## 1. INTRODUCTION

Virtual Observatory (VO) is a data intensive online astronomical research and education environment, taking advantages of advanced information technologies to achieve seamless, global access to astronomical information.

All sciences, including astronomy, are now entering the era of information abundance. The exponentially increasing volume and complexity of modern data sets promises to transform the scientific practice, but also poses a number of technological challenges. The Virtual Observatory concept is the astronomical community's response to these challenges (Djorgovski 2005). The power of the World Wide Web is its transparency. It is as if all the documents in the world are inside your PC. The idea of the Virtual Observatory is to achieve the same transparency for astronomical data and other related information (Quinn et al. 2004).

The VO concept was defined in late 1990's through many discussions and workshops, culminating in a significant endorsement by the U.S. National Academy's influential "decadal survey" report (McKee C., et al., 2000), and a white paper (U.S. NVO White Paper, 2001). The VO concept has a high priority in most national astronomy programs. In June 2002, at an international VO meeting in Munich, the International Virtual Observatory Alliance (IVOA) was formed (http://www.ivoa.net). The alliance of 16 member projects (Europe, US, UK, Canada, China, Russia, South Korea, Hungary, France, Germany, Italy, Australia, Japan, India, Spain, Armenia) seeks to ensure that the essential VO infrastructural technologies and interoperability standards are developed to enable a VO capability on a global scale.

The IVOA has created working groups and interest groups to pursue the discussion and definition of IVOA standards and mechanisms. To facilitate the global advancement of working group activities, the IVOA organizes Interoperability Workshops twice a year (Spring and Autumn) at which all working groups discuss progress and work towards final definitions of standards.

Interoperability is a basic and the most important requirement to achieve transparent access to global data and services. As the first IVOA Recommendation published in April 2002, VOTable (http://www.ivoa.net/Documents/latest/VOT.html) is an XML (eXtensible Markup Language, http://www.w3.org/XML/) standard for the interchange of data represented as a set of tables. VOTable was created from Astrores (http://cdsweb.u-strasbg.fr/doc/astrores.htx), an XML structure of tabular results. More details about VOTable history and chronology are available at "http://cdsweb.u-strasbg.fr/doc/VOTable/". At present, VOTable format has been widely adopted and supported by the VO community, data centers, observation projects, software applications and services (http://www.ivoa.net/twiki/bin/view/IVOA/VOTableSoftware).

Although VO will be a completely new web-based research environment, it is evolved from existing applications and services, which will be continually used by or integrated into the VO (Cui et al. 2004). OpenOffice.org (http://www.openoffice.org) is a mature, open source, front office applications suite with the advantage of native support OpenDocument XML file format (http://www.oasis-open.org/committees/tc_home.php?wg_abbrev=office). Taking advantage of eXtensible Stylesheet Language Transforms (XSLT), VOFilter (http://services.china-vo.org/vofilter/), an XML filter is developed by the Chinese Virtual Observatory (China-VO) project (Cui, Zhao et al. 2004). Using the filter, VOTable files can be loaded in OpenOffice.org Calc, a spreadsheet application, and analyzed as other spreadsheet files. Acting as a connector, the VOFilter bridges the coming VO and current industrial office applications.

In the paper, we first introduce some basic background about VOTable, OpenDocument, OpenOffice.org and its XML filter environment. In section 3, workflow of the VOFilter is presented. Installation and usage of the VOFilter are described in the fourth part. Finally, some problems and future development plans are discussed.

## 2. TECHNIQUE BACKGROUNDS

As for every software package, some technical backgrounds are required to learn how the VOFilter works. In the section, we give a brief introduction of IVOA VOTable, OpenDocument, OpenOffice.org and its XML filter environments.

### 2.1 IVOA VOTable

VOTable is the first Recommendation defined by the IVOA. VOTable 1.0 was released in April 2002, before the formation of the IVOA. The latest version of the Recommendation is 1.1, released on August 11,2004.

VOTable is an XML standard designed as a flexible storage and exchange format for tabular data, with particular emphasis on astronomical tables. Interoperability is encouraged through the use of XML standard family. In this context, a table is an unordered set of rows, each of a uniform format, as specified in the table metadata. Each row in a table is a sequence of table cells, and each of these contains either a primitive data type, or an array of such primitives.

An important feature of the format is the possibility of storing metadata and data separately, which makes the VOTable standard ready for the merging Grid computing (Foster, et al 2002). Data model of the VOTable is expressed in Table 1.

| | | |
|---:|:---|:---|
| **VOTable** | = | hierarchy of **Metadata** + associated **TableData**, arranged as a set of **Tables** |
| **Metadata** | = | **Parameters** + **Infos** + **Descriptions** + **Links** + **Fields** + **Groups** |
| **Table** | = | list of **Fields** + **TableData** |
| **TableData** | = | stream of **Rows** |
| **Row** | = | list of **Cells** |
| **Cell** | = | **Primitive** or variable-length list of **Primitives** or multidimensional array of **Primitives** |
| **Primitive** | = | integer, character, float, floatComplex, etc |

Table 1: VOTable data model

The data part in a VOTable may be represented using one of three different formats: TABLEDATA, FITS and BINARY. TABLEDATA is a pure XML format; FITS and BINARY are binary formats.

**2.2 OpenDocument**

OpenDocument, short for the OASIS Open Document Format for Office Applications, is an open document file format for saving and exchanging editable office documents. This standard is developed by OASIS (Organization for the Advancement of Structured Information Standards, http://www.oasis-open.org) consortium, based upon the XML-based file format originally created by OpenOffice.org (http://www.openoffice.org). The standard was publicly contributed by a variety of organizations, is publicly accessible, and can be implemented by anyone without any restriction.

The specification defines an XML schema for office applications and its semantics. The schema defines suitable XML structures for office documents and is friendly to transformations using XSLT or similar XML-based tools (Wheele 2005). In OpenDocument v1.0, six document types are defined: text, spreadsheet, chart, drawing, image and presentation.

Spreadsheet is mainly used for tabular data presentation and analysis. The content of spreadsheet documents mainly consists of a sequence of tables. Additionally, a spreadsheet document may contain forms, change tracking information and various kinds of declarations that simplify the usage of spreadsheet tables and their analysis. Structure of an OpenDocument Spreadsheet document is shown in Table 2.

```
<define name="office-body-content" combine="choice">
   <element name="office:spreadsheet">
      <ref name="office-spreadsheet-attlist"/>
      <ref name="office-spreadsheet-content-prelude"/>
      <ref name="office-spreadsheet-content-main"/>
      <ref name="office-spreadsheet-content-epilogue"/>
   </element>
</define>
```

Table 2: OpenDocument Spreadsheet document structure

The prelude contains the document's form data, change tracking information, calculation setting for formulas, validation rules for cell content and declarations for label ranges.

The main document is a list of tables, shown in Table 3. The structure of OpenDocument tables is similar to the structure of HTML4, XSL or VOTable tables.

```
<define name="office-spreadsheet-content-main">
  <zeroOrMore>
    <ref name="table-table"/>
  </zeroOrMore>
</define>
```

Table 3: Main document structure of OpenDocument Spreadsheet

The epilogue contains declarations for named expressions, database ranges, data pilot tables, consolidation operations and DDE (Dynamic Data Exchange) links.

### 2.3 OpenOffice.org

OpenOffice.org is an open source project mainly sponsored by Sun Microsystems (http://www.sun.com). OpenOffice.org, the product is an international multi-platform office suite. In OpenOffice.org 2.0 suite, six desktop applications are included: word processor (Writer), spreadsheet (Calc), formula editor (Math), drawing program (Draw), presentation program (Impress), database application (Base).

Beginning with version 2.0, OpenOffice.org uses the open standard OASIS OpenDocument XML format as its default file format. Organizations and individuals that store their data in an open format avoid being locked in to a single software vendor, leaving them free to switch software (Ogbuji, 2003). In addition to OpenOffice.org, another open source office suite KOffice (http://www.koffice.org) as well as OpenOffice.org derivatives like the StarOffice (http://www.sun.com/staroffice) support the OASIS OpenDocument file format. Furthermore, many communities like European Commission and US Massachusetts State, have adopted or are planning to adopt this open document format.

What make the VOFilter possible are the XML filter mechanisms provided by OpenOffice.org. An XML filter contains stylesheets that are written in the XSLT language. The stylesheets define the transformation between the OpenDocument file format and another XML format through export and import filters. In OpenOffice.org, there are three types of XML filters:

- Import Filters, load external XML files and transform the format of the files into the OpenDocument XML file format.
- Export Filters, transform OpenDocument XML files and save the files to a different XML format.
- Import/Export Filters, load and save OpenDocument XML files into a different XML format.

One can create one's own XML filter that convert a specified file format into the OpenDocument XML file format, or convert the OpenDocument XML file format into one's own file format. XML filter can be integrated into OpenOffice.org seamlessly. OpenOffice.org provides a friendly XML filter managing and packaging environment, "XML Filter Settings" dialog, where one can create, edit, delete, and test filter. With its help, one can save one's XML filter as a package and distribute it easily, just as the VOFilter do.

## 3. VOFilter WORKFLOW

VOFilter is an "import filter" to convert files from the VOTable XML format to the

OpenDocument Spreadsheet XML format. In OpenOffice.org 2, the suffix for spreadsheet documents is "ods", short for "OpenDocument Spreadsheet".

Following the structure of VOTable file introduced in section 2, three levels of transformation are performed during a converting process: "TABLE", "RESOURCE" and "VOTABLE". Several XSL templates are defined to complete the conversion. A top-level view of the VOFilter workflow is shown in Table 4.

```
<xsl:template name="vofilter">
    <office:automatic-styles>
        <xsl:call-template name="odsStyles"/>
    </office:automatic-styles>
    <office:body>
        <xsl:call-template name="mainConvert"/>
    </office:body>
</xsl:template>
<xsl:template name="mainConvert">
    <office:spreadsheet>
        <xsl:for-each select="//TABLE">
            <xsl:call-template name="cTABLE"/>
        </xsl:for-each>
        <xsl:for-each select="//RESOURCE">
            <xsl:call-template name="cRESOURCE"/>
        </xsl:for-each>
        <xsl:for-each select="/VOTABLE">
            <xsl:call-template name="cVOTABLE"/>
        </xsl:for-each>
    </office:spreadsheet>
</xsl:template>
<xsl:template name="cTABLE">
    <xsl:if test="//DATA/TABLEDATA">
        <xsl:call-template name="cTABLEDATA"/>
    </xsl:if>
    <xsl:if test=".//FIELD">
        <xsl:call-template name="cTABLEFIELD"/>
    </xsl:if>
    <xsl:if test=".//PARAM">
        <xsl:call-template name="cTABLEPARAM"/>
    </xsl:if>
    <xsl:if test=".//GROUP">
        <xsl:call-template name="cTABLEGROUP"/>
    </xsl:if>
    <xsl:call-template name="cTABLEother"/>
</xsl:template>
<xsl:template name="cRESOURCE">
    <xsl:call-template name="cRESOURCEPARAM"/>
    <xsl:call-template name="cRESOURCEINFO"/>
    <xsl:call-template name="cRESOURCEother"/>
</xsl:template>
<xsl:template name="cVOTABLE">
    <xsl:call-template name="cVOTABLEPARAM"/>
    <xsl:call-template name="cVOTABLEINFO"/>
    <xsl:call-template name="cVOTABLEother"/>
</xsl:template>
```

Table 4: VOFilter workflow

First, "odsStyles" template is called to build necessary OpenDocument Spreadsheet styles. "mainConvert" template is the core part of the filter to convert file content from the VOTable format to the "ods" format. In the main convert process, three levels of transforms are preformed. First, convert all "TABLE" elements, then all "RESOURCE" elements. In the end, information relating with the root element "VOTABLE" is converted.

During the "TABLE" transform, five possible parts are included, "DATA", "FIELD", "PARAM", "GROUP" and "other". These five parts cover most content under "TABLE" elements. "RESOURCE" and "VOTABLE" level transforms are simpler and similar, including three sections to complete conversion of "PARAM", "INFO" and other elements and attributes.

The transform process mentioned above covers every part of the VOTable schema, and is general enough to be used in other transform applications, like VOTable2XHTML (http://services.china-vo.org/votable2xhtml/).

Sheet name designation is an issue worth mentioning here. Usually, a VOTable file will be transformed into several or even tens of sheets by the VOFilter, depending on the structure of the file. Each sheet in a spreadsheet document must have a different name, i.e. sheet names are unique in an OpenOffice Calc file. To avoid collision, each sheet is designed a name using "ID" or "name" attribute of ELEMENT from which it is taken, and sheet content type. If neither "ID"s nor "name"s are available, words like "TABLE", "RESOURCE", "VOTABLE" will be used directly.

## 4. INSTALLATION AND USAGE

The VOFilter is packaged into a "jar" file, which is available at the website of the China-VO project. Installation process is very simple and is described in the following steps.

1. Start the OpenOffice.org suite and then create or open a Spreadsheet document.
2. Choose **Tools - XML filter settings**, shown as Fig. 1.

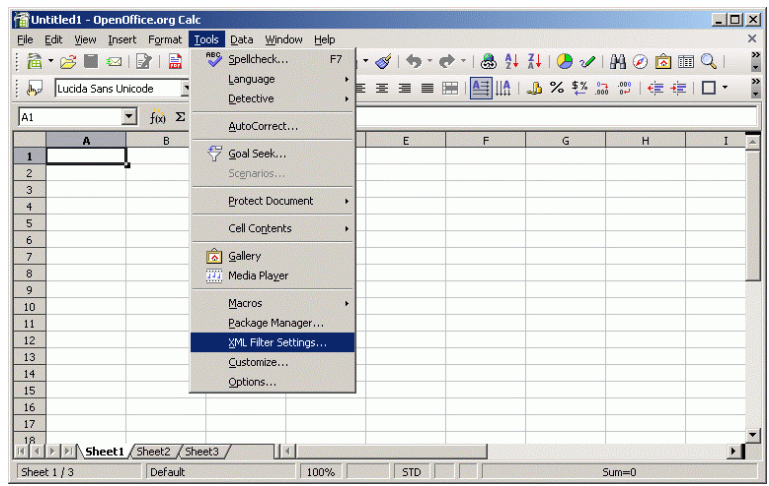

Fig. 1 OpenOffice.org Calc Interface

3. Click **Open Package** and select the package file that is downloaded from the website.

4. A message will be popped up telling the filter is installed successfully. "VOFilter" will be added into the Filter List, shown as Fig. 2.

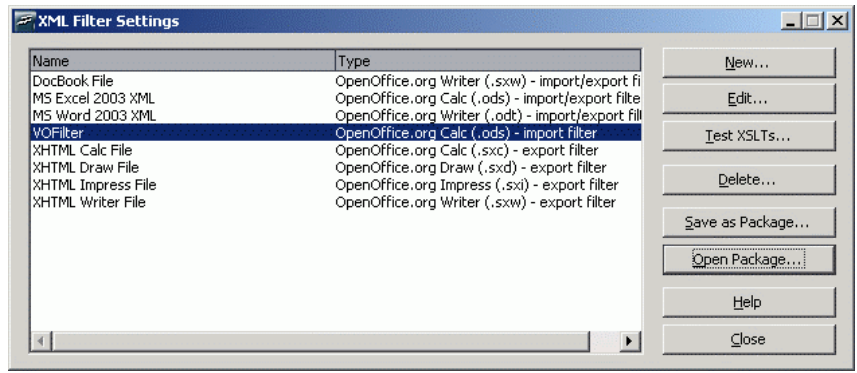

Fig. 2   "XML Filter Settings" dialog of OpenOffice.org Calc

5. "IVOA VOTable" option will appear in "**Open**" file type list. Now, VOTable files can be loaded and analyzed as native Spreadsheet files. In order to let users test the filter, a sample VOTable file is provided at the website, which is based on the output of a VizieR (http://vizier.u-strasbg.fr/) query on GSC 2.2.1 catalog.

## 5. DISCUSSION AND CONCLUSION

VOTable data files are organized in a three-dimension tree scheme. While OpenDocument spreadsheet files are organized in two-dimensional plain topology. In order to keep all relations among VOTable elements and attributes, many additionally columns and sheets must be added during the transform from VOTable structure to OpenDocument structure. To keep the output spreadsheet clean and not over-redundant, the VOFilter abandons some components of the source VOTable file during transform.

In the present version (2.0), non-supported elements and attributes, some problems and limitations, are listed as follows.

- Only "TABLEDATA" data content is supported. "FITS" and "BINARY" serializations are NOT supported.
- VOTable elements and attributes not supported:
    - Recursive "RESOURCE" elements
    - Recursive "GROUP" elements
    - "VALUES" elements for "FIELD" and "PARAM" elements
    - All attributes except "href" of "LINK" elements
    - "encoding" attribute of "TD" elements
- To avoid information and precision loss, all types of data in VOTable files are loaded as "text".
- For a VOTable file including multiple TABLEs without any "ID" or "name" attributes for "VOTABLE", "RESOURCE", and "TABLE" elements, the filter will crash for sheet name collision in OpenOffice.org 1.9.118 and earlier versions. In OpenOffice 1.9.122 and later versions, sheets will be renamed as "Table*1*" to "Table*n*" automatically by the office suite.
- Maximum row number, column number and sheet number are limited by the OpenOffice.org.
- XSLT transform is a memory consuming process. Transformations on very large VOTable files, for example larger then 200MB, may be questionable.

Some of above problems and limitations will be fixed in future versions. However, some are intrinsic ones of the VOTable, XML, OpenDocument and the OpenOffice.org, which are out of scope of the VOFilter. The non-supported elements and attributes mentioned above are mainly recursive or deep level ones. To support them, additional sheets and columns must be added into the output, which will make the spreadsheet very redundant.

Limited by functions of XSLT and OpenOffice.org XML Filter mechanism, it is very hard for VOFilter to support "FITS" and "BINARY" serializations directly. Thanks to the wide support of VOTable, there is an indirectly solution. First, convert "FITS" or "BINARY" serializations into "TABLEDATA" by third-party tools, for example TOPCAT (Taylor, 2005). Then transform it into OpenDucument format using the VOFilter.

In an earlier version of the VOFilter, i.e. 1.0, an export filter was included, while in version 2.0, it is temporarily deprecated. An important concern about the export filter is redundancies in the output VOTable file: a large fraction of the VOTable files available now are trimmed ones, usually created on purpose by programs and on-line services from various data centers and projects, that keep only the necessary parts. To ensure that the output is fully compatible with its XML schema, the export filter has to patch the original VOTable file with all the elements and attributes described in the VOTable XML schema. Another reason of deprecating the export filter is that, learned from discussions with some astronomers and VO developers, the operation of saving back a spreadsheet into a VOTable file is uncommon. If there is an obvious requirement for an export filter, it will be reintroduced in a future version.

VOFilter, the XML filter to convert VOTable format into OpenDocument Spreadsheet format, bridges the merging VO and industrial office applications. This will encourage the adoption of the VOTable format, and furthermore adoption of many other VO services.

Taking the power of opening, OpenDocument XML file format is obtaining wider and wider adopting and support. Additional to OpenOffice.org, in the future versions of the VOFilter, it will adapt to wider running environments where OpenDocument XML file format is supported.

Both VOTable and OpenDocument are specifications under evolution. VOFilter will also be upgraded periodically to keep compatibility with the latest specifications and further requirements.

**Acknowledgements**

Bruno Rino, European Southern Observatory, is acknowledged for his detail test of the VOFilter and comprehensive feedback. Cui C. thanks Jayant Gupchup, John Hopkins University, for his carefully check of the manuscript. Cui C. also thanks André Schaaff and Mark Allen for fruitful discussion. Cui C. thanks Observatoire Astronomique de Strasbourg, and Université Louis Pasteur, for financial support during a stay to the CDS, during which part of this work was completed. Cui C. thanks the China-VO team for their collaboration. The authors would like to thank referee François Ochsenbein for valuable comments and suggestions that improved the article. In particular, special thanks are due to Kui WU, who developed the base code of the first version of the VOFilter. The China-VO project is funded by the National Natural Science Foundation of China under contract 90412016.